\documentclass[twocolumn,superscriptaddress,showpacs,amsmath,amssymb]{revtex4}
\usepackage{graphicx}
\usepackage{amssymb}
\usepackage{amsmath}
\usepackage{dcolumn}
\usepackage{bm}

\begin{document}
\preprint{}

\title{Epidemic Dynamics of Interacting Two Particle Species on 
       Scale-free Networks}

\author{Yong-Yeol Ahn}
\affiliation{Department of Physics, Korea Advanced Institute of Science and Technology, 
Daejeon 305-701, Korea}
\author{Naoki Masuda}
\affiliation{Amari Research Unit, RIKEN Brain
  Science Institute, 2-1, Hirosawa, Wako, Saitama 351-0198, Japan}
\author{Hawoong Jeong}
\affiliation{Department of Physics, Korea Advanced Institute of Science and Technology, 
Daejeon 305-701, Korea}
\author{Jae Dong Noh}
\affiliation{Department of Physics, University of Seoul,
  Seoul 130-743, Korea}
\date{\today}
\begin{abstract}
We study the non-equilibrium phase transition in a model for epidemic 
spreading on scale-free networks.
The model consists of two particle species $A$ and $B$, and
the coupling between them is taken to be 
asymmetric; $A$ induces $B$ while $B$ suppresses $A$. 
This model describes the spreading of an epidemic on networks 
equipped with a reactive immune system.
We present analytic results on the phase diagram and the critical
behavior, which depends on the degree exponent $\gamma$ of the underlying
scale-free networks. Numerical simulation results that support the analytic
results are also presented.

\end{abstract}

\maketitle

Network concept has been emerging as a useful tool for the study  of complex
interconnected systems~\cite{albert02, dorogovtsev02, newman03}.  It
allows us to study structure and dynamics of those systems in various
disciplines in a simple unified manner.  Many networks in nature
share an intriguing property of the power-law degree distribution
$P(k) \sim k^{-\gamma}$ with the degree exponent $\gamma$, where
$P(k)$ is the probability of a node having $k$ links.  A network with
this property is called a scale-free~(SF) network~\cite{albert02},
examples of which include the world-wide web~\cite{albert99}, the
Internet~\cite{faloutsos99}, and the human sexual contact
network~\cite{liljeros01}, and so on.

The broad degree distribution of the SF network brings about various
interesting phenomena. Particularly we are interested in epidemic
spreading. Networked systems are susceptible to epidemics, which may
be computer viruses in computer networks or infectious diseases in
social contact networks of individuals.  Robustness against
the epidemic spreading is one of the key elements for a proper
function of networks.  Researchers have studied the characteristic
feature of the epidemic spreading in the SF network and the efficient
immunization strategy suitable for the SF network~\cite{satorras01,
  boguna03, satorras02, dezso02, cohen03,newman05}.

There are hubs with a large number of links in the SF network.  This
structure makes it vulnerable to an epidemic since it can spread easily
through those hubs.  It is found that an epidemic even with an
extremely small spreading rate never ceases spreading in the SF networks
with $\gamma\leq 3$~\cite{satorras01, boguna03}.
In those networks, the random immunization where all nodes are
immunized or vaccinated with an equal but finite probability becomes 
inefficient.  Instead, the targeted immunization where hubs are
immunized preferentially proves to be 
efficient~\cite{satorras02, dezso02, cohen03}.

In the study of the epidemic spreading, one usually considers the dynamics
of malicious agents on bare or immunized networks. On the other hand, the
following examples show that the competing spreading dynamics between malicious
agents and immunizing agents are also important.
Recently, the worm ``Code Red'' almost paralyzed the whole Internet by 
flooding it with lots of useless packets~\cite{moore02}.  
After the outbreak, there appeared the so-called worm-killer worm 
``Code Green'' which was meant 
to seek out and kill the malicious worm~\cite{sans}.
Although it also flooded the network and did more harm than good,
it hinted a possibility of contagious vaccination.
The competing dynamics is also observed inside living organisms. 
When pathogens invade and spread, immune cells are stimulated. 
The awaken immune cells then replicate themselves and get rid of the 
pathogens~\cite{coico}. 
The phenomenon of {\em
cross immunity} between competing pathogens is another
example~\cite{newman05}. If populations are exposed to a certain disease,
then they become immune to the other for certain pairs of diseases, 
e.g., Hansen's disease and tuberculosis~\cite{karlen}.

In this work, we introduce a model that mimics the competing spreading
dynamics and study the phase transition it displays on SF networks.
The model consists of two species~($A$ and $B$) particles, one of which
infects nodes and the other of which heals the infection. 
They may represent worms~($A$)
and worm-killer worms~($B$) in computer networks or malicious
pathogens~($A$) and immune cells~($B$) in living organisms.

At each time step particles evolve according to the following dynamic rule:
A particle $A$ annihilates spontaneously with the probability $p_A$. Or,
with the probability $q_A= 1-p_A$, it branches offsprings to its all
neighboring
nodes, a fraction $1-\lambda$~($\lambda$) of which is the species $A~(B)$.
A particle $B$ annihilates spontaneously with the probability $p_B$.
Or it branches an offspring $B$ to one of its neighboring nodes with the
probability $q_B=1-p_B$. A particle $A$ is removed with the probability
$\mu$ upon contact with a particle $B$ at the same node.

Each node may be empty, occupied by a particle $A$, or 
occupied by a particle $B$. Such a node can be interpreted as a healthy or
susceptible, infected, or immunized individual, respectively. 
Then the process with $p_A$ and $q_A$
corresponds to spontaneous healing and infection, respectively. The
process with $\lambda$ corresponds to activation of an immune system. 
The spontaneous annihilation of $B$ particles takes account of the fact that
immunization may not be permanent. 
The parameter $\mu$ describes the efficiency of the immune
system in healing the infection. 

Note that the particles $A$ and $B$ have different branching rules.
A particle $A$ branches offsprings to {\em all} neighboring nodes as 
in the susceptible-infected-susceptible~(SIS) 
model~\cite{satorras01}, while a particle $B$ branches an offspring 
to {\em one} of the neighboring nodes as in the contact
process~(CP)~\cite{castellano06}. 
This difference can be negligible in networks with the narrow degree 
distribution, but it makes a big difference in SF networks~\cite{castellano06}. 
We adopt the less-reproductive CP dynamics for the particle $B$ since
it would be costly to activate an immune system in real systems. 
The model was studied in networks with a narrow degree 
distribution~\cite{noh05}. As we will see, the model shows much richer
properties in SF networks.

The model displays three stationary state phases denoted by 
$(0,0)$, $(0,B)$, and $(A,B)$. They are distinguished with the 
stationary state density of each particle species. 
Both species are {\em inactive} 
with zero density in the $(0,0)$ phase. In the $(0,B)$ phase, 
the $A$ particles are inactive while
the $B$ particles are {\em active} with a nonzero density. 
Both species are active with nonzero densities in
the $(A,B)$ phase. Due to the asymmetric nature of the interaction, one does
not have a phase in which the $A$ particles are active and
the $B$ particles are inactive.
All nodes are healthy in the $(0,0)$ and $(0,B)$ phases. The healthy state
requires that the $B$ particles are active in the $(0,B)$ phase, 
which is not the case in the $(0,0)$ phase. 
In this sense, the $(0,0)$ phase is the ideal one since 
the remnant $B$ particles would cost system resources.

We will study the phase transitions and the critical behaviors 
in SF networks by adopting the rate equation approach
which proves useful for a single species or multi species particle
systems~\cite{satorras01,masuda05}.
It is assumed that the particle density at a node is given
by a function of its degree only. Let $a_k$ and $b_k$ be the density of $A$
and $B$ particles on nodes with degree $k$, respectively. Then, it is
straightforward to show that they satisfy the coupled rate equations
\begin{equation}
\begin{split}
  \dot{a}_{k} &= -p_{A} a_{k} + q_A (1-\lambda) k (1-a_{k}) \Theta_{A}(k) -
                             \mu k a_{k} \Theta_{B}(k)\\
  \dot{b}_{k} &= -p_{B} b_{k} + q_B k (1-b_{k}) \Theta_{B}(k) 
    + \lambda k (1-b_{k}) \Theta_{A} (k),
\end{split}
\end{equation}
where $\Theta_{A}(k)$~($\Theta_{B}(k)$) denotes 
the probability that a
node with degree $k$ is infected by a particle $A$~($B$) from each of its 
neighboring
nodes. They are given by $\Theta_A(k) = \sum_{k'} a_{k'} P(k'|k)$ 
and $\Theta_B(k) = \sum_{k'} b_{k'} P(k'|k)/k'$, respectively. Here $P(k'|k)$
is the conditional probability that a node at one end of a link 
have the degree $k'$ under the condition that the other end of the link
has the degree $k$. The different form of $\Theta_A$ and $\Theta_B$ is due
to the different spreading dynamics of the two particle species.
The conditional probability can take care of a degree 
correlation~\cite{newman02}. In
this work, we only consider networks with no degree correlation, that is,
$P(k'|k) = k' P(k') / \langle k \rangle $ with the degree distribution
$P(k) \sim k^{-\gamma}$ and the mean degree $\langle k
\rangle$~\cite{newman02}. Then, one obtains that
\begin{equation}
\Theta_{A} = \frac{1}{\langle k\rangle} \sum_{k} k\, a_{k} P(k) ,
\quad \Theta_{B} = \frac{1}{\langle k\rangle} \sum_{k} b_{k} P(k).
\end{equation}
They are related to the particle density and 
will be used as the order parameter.

In the stationary state~($\dot{a}_{k} = \dot{b}_{k} = 0$), the order
parameter satisfies the coupled self-consistency equations 
\begin{subequations}\label{self}
\begin{align}
\label{selfa} \Theta_{A} &= f(\Theta_A,\Theta_B)\equiv  \frac{1}{\langle k \rangle} \sum_{k} 
\frac{\tilde{q}_A \Theta_{A} k^{2} P(k) }{1+(\tilde{q}_A \Theta_{A} +
\tilde{\mu} \Theta_{B}) k},\\
\label {selfb} \Theta_{B} &= g(\Theta_A,\Theta_B)\equiv  \frac{1}{\langle k \rangle} \sum_{k} 
\frac{(\tilde{\lambda} \Theta_{A} + \tilde{q}_B \Theta_{B}) k P(k)
}{1+(\tilde{\lambda} \Theta_{A} + \tilde{q}_B \Theta_{B}) k}, 
\end{align}
\end{subequations}
where $\tilde{q}_A \equiv \frac{q_{A}}{p_{A}}(1-\lambda )$, $\tilde{\mu} 
\equiv \frac{\mu}{p_{A}}$, $\tilde{\lambda} \equiv \frac{q_{A}}{p_{B}} 
\lambda$ , and $\tilde{q}_B \equiv \frac{q_{B}}{p_{B}} $.  
Note that, without the interaction~($\tilde{\lambda}=\tilde{\mu}=0$), 
the self-consistency equations are decoupled, each of which has been studied
separately~\cite{satorras02,castellano06}.

\begin{figure}
 \includegraphics[width=0.9\columnwidth]{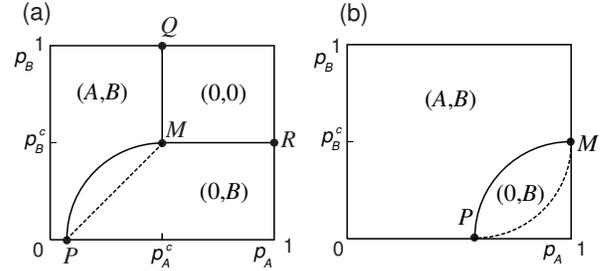}
 \caption{The schematic phase diagram in SF
   networks with $\gamma >  3$ in (a) and $\gamma \leq 3$  in (b). 
   The phase boundary $\overline{MP}$ changes its shape as $\gamma$ varies.
   In (a), it is tangential to $\overline{MR}$ for $\gamma<4$ (solid
line) and non-tangential to $\overline{MR}$ or $\overline{MQ}$ for
$\gamma\geq4$~(dashed line). In (b), it is parallel to $p_A$ axis at $M$ for
$\gamma>5/2$~(solid line), and to $p_B$ axis near $M$ for
$\gamma<5/2$~(dashed line).}
  \label{scheme}
\end{figure}

The self-consistency equations allow us to determine the phase diagram,
whose schematic plot is shown in Fig.~\ref{scheme}. Let us summarize the
result first: For $\gamma>3$~(Fig.~\ref{scheme}(a)), 
we find that the system exhibits the three phases 
$(0,0)$, $(0,B)$, and $(A,B)$. 
Across the line $\overline{MR}$, $A$ particles remain inactive 
and $B$ particles become active. 
Across the line $\overline{QM}$, both particles 
become active simultaneously. 
Across the line $\overline{MP}$,  $B$ particles are already active and
the $A$ particles become active. The three phase transition
lines merge into the multicritical point $M$.

On the other hand, the system exhibits the
only two phases $(0,B)$ and $(A,B)$ for $\gamma\leq
3$~(Fig.~\ref{scheme}(b)). Without the $B$ species, the $A$ species would
always be in the active phase for $\gamma\leq 3$~\cite{satorras01}. 
The existence of the phase $(0,B)$ implies
that one can prevent an epidemic from spreading 
even in the SF networks with $\gamma\leq 3$. 
The nonexistence of the phase $(0,0)$, however, implies 
that the immunization is possible only when the species $B$ 
is kept to be in the active state, which may cost system resources.

Now we sketch briefly the way the phase diagram is obtained, details of
which will be presented elsewhere~\cite{unpub}.
It is obvious that $\Theta_A=\Theta_B=0$ is a solution of the
self-consistency
equation. One can obtain the boundary of the phase $(0,0)$
from the condition for the existence of nonzero solutions for $\Theta_A$ or
$\Theta_B$.
From Eq.~\eqref{selfa}, we find that $f(0,0)=0$ and $f(\Theta_A,0)\leq 1$
and $\partial^2f(\Theta_A,0)/\partial \Theta_A^2<0$ for all $\Theta_A$. 
So a nonzero solution for $\Theta_A$ exists 
when $\partial f(0,0)/\partial \Theta_A =
\tilde{q}_A \langle k^2 \rangle / \langle k \rangle \geq 1$ with $\langle
k^2 \rangle$ the second moment of the degree. This gives the phase
boundary $\overline{QM}$ at $\tilde{q}_A=\langle k\rangle / \langle
k^2\rangle$ or at $p_A = p_A^c$ with
\begin{equation}
p_{A}^{c} = \frac{(1-\lambda)  {\langle k^{2} \rangle}/{\langle k \rangle}}
{1 + (1-\lambda) {\langle k^{2} \rangle}/{\langle k \rangle}} \ .
\label{pAc}
\end{equation}
Similarly, a nonzero solution for $\Theta_B$ exists 
when $\partial g(0,0)/\partial \Theta_B = \tilde{q}_B \geq
1$, which gives the phase boundary $\overline{MR}$ at $\tilde{q}_B=1$ or
at $p_B = p_B^c=1/2$.
For $\gamma\leq 3$, $\langle k^2 \rangle$ is infinite, $p_A^c=1$, and
the phase $(0,0)$ vanishes.

At the phase boundary $\overline{MP}$, $\Theta_A=0$ and $\Theta_B$ has a
nonzero value $\Theta_B^\star$ satisfying $\Theta_B^\star = 
g(0,\Theta_B^\star)$. 
We also have that $\partial f(0,\Theta_B^\star)/\partial \Theta_A = 1$ 
since $\Theta_A$ begins to deviate from zero at that line. 
These two equations define the phase boundary $\overline{MP}$. 
Near the point $M$, the phase boundary is given by
\begin{equation}
\epsilon_B \sim 
\begin{cases}
\epsilon_A^{\ 1/\mbox{min}(1,\gamma -3)}, & \gamma > 3 , \\
\epsilon_A^{(\gamma -2)/(3-\gamma)}, & \gamma < 3  ,
\end{cases}
\end{equation}
for small $\epsilon_A = p_A^c - p_A$ and $\epsilon_B =p_B^c -
p_B$~\cite{unpub}.

We also studied the critical behavior of the order parameter near the phase
transitions by analyzing the property of 
the functions $f$ and $g$ at small values of $\Theta_A$ and $\Theta_B$. 
We found that the order parameter shows the power-law scaling 
\begin{equation}
\Theta_{A,B} \sim \epsilon^{\beta_{A,B}}
\end{equation}
with the $\gamma$ dependent critical exponents $\beta_A$ and $\beta_B$ for
$\gamma\neq 3$. The critical exponents have different values at different
critical lines. The results are summarized in Table~\ref{beta}. The model
displays interesting multicritical behaviors at the multicritical point $M$,
which will be discussed elsewhere~\cite{unpub}.

It is noteworthy that the model displays the peculiar critical behaviors at
$\gamma=3$.
Let us take the degree distribution as
$P(k) = c k^{-3}$ for $k_0 \leq k$ with a
normalization constant $c$ and a degree cutoff $k_0$. 
With the continuum $k$ approximation in Eq.~(\ref{self}), 
the functions $f$ and $g$ are given by
$f =  \frac{c \tilde{q}_A\Theta_A }{\langle k \rangle} \ln (\frac{1+k_0
X}{k_0 X})$
and $ g =  Y - \frac{c}{\langle k\rangle} Y^2 \ln (\frac{1+k_0 Y}{k_0 Y})$ 
with $X\equiv \tilde{q}_A \Theta_A + \tilde{\mu}\Theta_B$ and 
$Y \equiv \tilde\lambda\Theta_A + \tilde{q}_B \Theta_B$.
The logarithmic dependence leads to the following peculiar 
behaviors~\cite{unpub}: 
Let $\epsilon_A = 1-p_A$ and $\epsilon_B = 1/2-p_B$ be the deviation from
$M$ at $p_A = 1$ and $p_B = 1/2$. 
The phase boundary $\overline{MP}$ is given by
\begin{equation}
\epsilon_B \sim \frac{1}{\epsilon_A} e^{-d / \epsilon_A} 
\end{equation}
with a constant $d$ for small $\epsilon_A$ and $\epsilon_B$. 
When one approaches $M$ from the phase $(A,B)$, the order parameter exhibits
a path dependent critical behavior. Along the path with $\epsilon_B=0$ that
is tangential to $\overline{MP}$, we find that 
\begin{equation}
\Theta_A \sim  \frac{1}{\epsilon_A^2} e^{-2 d / \epsilon_A} \quad ,  \quad
\Theta_B \sim e^{-d/\epsilon_A} .
\end{equation}
Along non-tangential paths with finite $\epsilon_A/|\epsilon_B|$,
we find that
\begin{equation}
\Theta_A \sim \Theta_B \sim e^{-d /\epsilon_A} \ .
\end{equation}
In contrast to the power-law scaling at $\gamma\neq 3$, 
the order parameter as well as the phase boundary 
has the essential singularity at $\gamma=3$. The order parameter also has
the essential singularity near the line with $p_A=1$ and $p_B>p_B^c$~(see
Table~\ref{beta}).
That is, the transition is the infinite order transition. 
The infinite order transition was reported recently in the percolation
problem and the equilibrium Ising model on growing
networks~\cite{bauer05,percolation}.

\begin{table}
\begin{center}
\begin{tabular}{l|c|c|c|c}
\hline\hline
 & & $(0,0) \to$ & $(0,0) \to$ & $(0,B) \to$ \\
 & & $(A,B)$ & $(0,B)$ & $(A,B)$ \\
\hline
$\gamma > 4$ & $\beta_A$ & 1 &   &  1  \\
             & $\beta_B$ & 1 & 1 &     \\
\hline
$3<\gamma <4$& $\beta_A$ & $\frac{1}{\gamma -3}$ & & 1 \\ 
             & $\beta_B$ & $\frac{1}{\gamma -3}$ & 1 &  \\ 
\hline
$2<\gamma < 3 $  & $\beta_A$ & $\frac{\gamma -2}{3-\gamma}$ &  &1 \\
             & $\beta_B$ & $\frac{1}{3-\gamma}$ & $\frac{1}{\gamma -2}$ & \\
\hline
$\gamma =3 $ & $\Theta_A$ & $\sim e^{-d/\epsilon_A}/\epsilon_A$ &  & 1 \\
             & $\Theta_B$ & $\sim e^{-d/\epsilon_A}$ & 1 & \\
\hline\hline
\end{tabular}
\end{center}
\caption{The exponents $\beta_A$ and $\beta_B$ associated with 
each phase transition line. For $\gamma\leq 3$, the phase $(0,0)$ denotes
the line with $p_A=1$ and $p_B>p_B^c$.}
\label{beta}
\end{table}

In order to confirm the phase diagram, we performed numerical simulations
on SF networks generated by using the so-called static model~\cite{kigoh01}.
Initially networks are filled with the particles,
and the density $\rho_A$ and $\rho_B$ of each particle species
is measured and averaged over at least 2000 simulations and 10
network realizations. All simulations were performed with 
$\lambda=\mu=0.5$. All numerical data presented below were obtained on the
static model networks with $N=64 \times 10^4$ nodes and the mean degree
$\langle k\rangle = 12$.

\begin{figure}
 \includegraphics[width=\columnwidth]{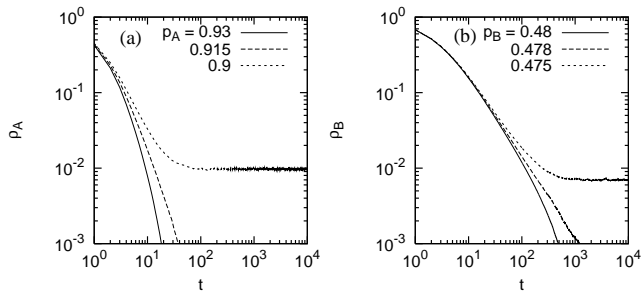}
 \caption{Density decay at $\gamma=3.5$. In (a) $\rho_A$ is plotted at 
$p_B=0.9$, and in (b) $\rho_B$ is plotted at $p_A=0.99$.}
 \label{fig2}
\end{figure}

We present the numerical data from the SF networks with 
$\gamma=3.5$ in Figs.~\ref{fig2}~(a) and (b).
They show the signature of the phase transition from $(0,0)$ to $(A,B)$
at $p_A^c \simeq 0.915$ and $(0,0)$ to $(0,B)$ at $p_B^c \simeq 0.478$, 
respectively. These numerical data confirms the existence of the three 
phases for $\gamma>3$.

\begin{figure}
\includegraphics[width=0.9\columnwidth]{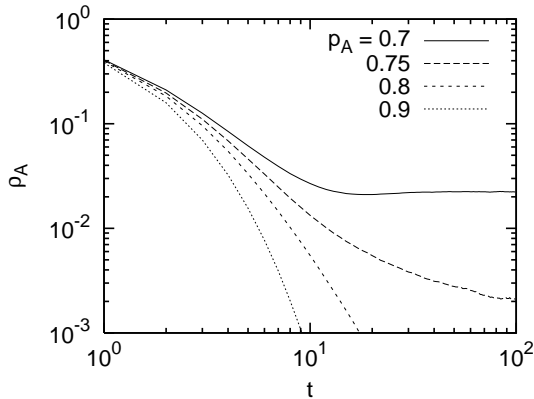}
 \caption{Density decay on SF networks with $\gamma=2.75$. $\rho_A$ is
plotted for several values of $p_A$ with $p_B=0.3$.}
 \label{fig3}
\end{figure}

At $\gamma=2.75$, we performed the simulations 
at $p_B=0.3$ and for several values of $p_A$ in order to examine the
existence of the $(0,B)$ phase. The numerical data presented in
Fig.~\ref{fig3} clearly show that the $A$ species is in the active phase at
$p_A=0.7$ and in the inactive phase at $p_A=0.9$. Although it is hard to
locate the critical point accurately, the numerical data supports the
analytic prediction that the $A$ species can be inactive even for
$\gamma\leq 3$ in our model.
At $\gamma=2.75$, we also performed the simulations at $p_B=0.9$ and at
several values of $p_A$ close to $1$ in order to examine whether the phase
$(0,0)$ exists or not. Numerical data obtained on the networks of
sizes up to $N=64\times10^4$ nodes indicate that the $A$ species is in the
active phase at least up to $p_A=0.99$ and that the $A$ species become more
active as $N$ increases~\cite{unpub}. 
So we conclude that the phase $(0,0)$ does not
exist in the asymptotic $N\to \infty$ limit.

In summary, 
we have studied the two-species epidemic model on SF networks. The two
species $A$ and $B$ are coupled asymmetrically in that the former induces
the latter whereas the latter suppresses the former. Our model is aimed at
describing the spreading dynamics of competing malicious 
pathogens~($A$ particles) and reactive immunizing agents~($B$ particles) 
on complex SF networks. The model is shown to have the different phase diagram
depending on whether $\gamma>3$ or $\gamma\leq 3$~(see Fig.~\ref{scheme}).
Our results show that one can prevent the epidemic from prevailing even in SF
networks with $\gamma\leq 3$. The results also show that it requires that
the immunizing agents should be kept in the active phase for $\gamma\leq 3$,
which is not necessary for $\gamma>3$. We have also investigated the critical
behaviors associated with the phase transitions. 
Especially, when $\gamma=3$, the phase transitions are infinite order
transitions with the essential singularity in the order parameters.

Acknowledgement: This work was supported by Korea
Research Foundation Grant (KRF-2004-041-C00139). YYA and HJ were supported
by the Ministry of Science and Technology through Korean Systems Biology
Research Grant (M10309020000-03B5002-00000). NM thanks the Special
Postdoctoral Researchers Program of RIKEN.


\begin{thebibliography}{50}
\bibitem{albert02} R. Albert and A.-L Barab\'asi, Rev. Mod. Phys.
\textbf{74}, 47 (2002).

\bibitem{dorogovtsev02} S. N. Dorogovtsev and J. F. F. Mendes,
Evolution of Networks, Advances in Physics \textbf{51}, 1079 (2002).

\bibitem{newman03} M. E. J. Newman,
SIAM Review \textbf{45}, 167 (2003).

\bibitem{albert99} R. Albert and H. Jeong and A.-L Barab\'asi, Nature
  (London) \textbf{401}, 130 (1999).
  
\bibitem{faloutsos99} M. Faloutsos and P. Faloutsos and C. Faloutsos,
  Computer Communications Rev. \textbf{29}, 251 (1999).
  
\bibitem{liljeros01} F. Liljeros and C. R. Edling and L. A. N. Amaral
  and H. E. Stanley and Y. Aberg, Nature (London) \textbf{411}, 907
  (2001).
  
\bibitem{satorras01} R. Pastor-Satorras and A. Vespignani, Phys. Rev.
  Lett. \textbf{86}, 3200 (2001).

\bibitem{boguna03} M. Bogu\~n\'a, R. Pastor-Satorras and A. Vespignani,
  Phys. Rev. Lett. \textbf{90}, 028701 (2003).

\bibitem{satorras02} R. Pastor-Satorras and A. Vespignani,
Phys. Rev. E \textbf{65}, 036104 (2002). 

\bibitem{dezso02} Z. Dezs\"o and A.-L. Barab\'asi, Phys. Rev. E
  \textbf{65}, 055103(R) (2002).
  
\bibitem{cohen03} R. Cohen and S. Havlin and D. ben-Avraham, Phys.
  Rev. Lett. \textbf{91}, 247901 (2003).
  
\bibitem{newman05} M. E. J. Newman, Phys. Rev. Lett. \textbf{95},
  108701 (2005).

\bibitem{moore02} D. Moore and C. Shannon and k claffy,
  \textit{Proceedings of the 2nd ACM SIGCOMM Workshop on Internet
    Measurement} (ACM Press, New York, NY, 2002), pp. 273-284.
  
\bibitem{sans} See the malware FAQ on Code Red worm 
  at http://www.sans.org/resources/malwarefaq/code-red.php
  in the SANS institute.

\bibitem{coico} R. Coico and G. Sunshine and E. Benjamini,
  \textit{Immunology: A short course}, (Wiley-Liss, 2003), 5th Ed. 

\bibitem{karlen} A. Karlen, \textit{Man and Microbes}, (Simon \&
  Schuster, 1996), 1st Ed. 

\bibitem{castellano06} C. Castellano and R. Pastor-Satorras,
Phys. Rev. Lett. \textbf{96}, 038701 (2006). 

\bibitem{noh05} J. D. Noh and H. Park, Phys. Rev. Lett. 
\textbf{94}, 145702 (2005).


\bibitem{masuda05} N. Masuda and N. Konno, J. Theor. Biol. (in press).

\bibitem{newman02} M. E. J. Newman, Phys. Rev. Lett. {\bf 89},
  208701 (2002).

\bibitem{unpub} Y.-Y. Ahn, H. Jeong, and J. D. Noh, unpublished.

\bibitem{percolation} 
D.S. Callaway, J.E. Hopcroft, J.M. Kleinberg, M.E.J. Newman, S.H. Stogatz, 
Phys. Rev. E {\bf 64}, 041902 (2001);
S.N. Dorogovtsev, J.F.F. Mendes, and A.N. Samukhin,
Phys. Rev. E {\bf 64}, 066110 (2001). 

\bibitem{bauer05} M. Bauer, S. Coulomb, and S. N. Dorogovtsev, 
        Phys. Rev. Lett. {\bf 94}, 200602 (2005).

\bibitem{kigoh01} K.-I. Goh and B. Kahng and D. Kim, 
  Phys. Rev. Lett. {\bf 87}, 278701 (2001). 
\end{thebibliography}
\end{document}